\documentstyle[12pt]{article}
\def\beq{\begin{equation}}
\def\eeq{\end{equation}}
\def\bea{\begin{eqnarray}}
\def\eea{\end{eqnarray}}

\def\ba{\begin{array}}
\def\ea{\end{array}}
\def\ds{\displaystyle}

\def\,{\"{U}}
\def\6{\.{I}}

\begin{document}
\baselineskip 0.7cm

\title{{\Large{Supersymmetric Solution of PT-/Non-PT-Symmetric and
Non-Hermitian Morse Potential via Hamiltonian Hierarchy Method}}}

\author{Metin Akta\c{s} and Ramazan Sever$\thanks{Corresponding author:
sever@metu.edu.tr}$\\
Department of Physics, Middle East Technical University \\
06531 Ankara, Turkey }

\date{\today}

\maketitle

\begin{abstract}

\noindent Supersymmetric solution of PT-/non-PT-symmetric and
non-Hermitian Morse potential is studied to get real and
complex-valued energy eigenvalues and corresponding wave
functions. Hamiltonian Hierarchy method is used in the calculations.\\

\end{abstract}

\newpage
\section{Introduction}
\noindent PT-symmetric Hamiltonians has acquired much interest in
recent years[1, 2, 3]. Bender and Boettcher [1] suggested that a
non-Hermitian complex potential with the characteristic of
PT-invariance has real energy eigenvalue if PT-symmetry is not
spontaneously broken. The other concept for a class of
non-Hermitian Hamiltonians is pseudo-Hermiticity. This kind of
operators satisfy the similarity transformation
$\eta~\hat{H}~\eta^{-1}=\hat{H}^{\dag}$ [3, 4, 5]. PT-invariant
operators have been analysed for real and complex spectra by using
a variety of techniques such as variational methods [7], numerical
approaches [8], semiclassical estimates [9], Fourier analysis [10]
and group theoretical approach with the Lie algebra [11]. It is
pointed out that PT-invariant complex-valued operators may have
real or complex energy eigenvalues [12]. Many authors have studied
on PT-symmetric and non-PT-symmetric non-Hermitian potential cases
such as flat and step potentials with the framework of SUSYQM
[13-15], exponential type potentials [16-21], quasi exactly
solvable quartic potentials [22-24], complex H\'{e}non-Heiles
potential [25], and therein [26-28]. In the present work, the real
and complex-valued bound-state energies of the q-deformed Morse
potential are evaluated through the Hamiltonian Hierarchy method
[29] by following the framework of PT-symmetric quantum mechanics.
This method also known as the factorization method of the
Hamiltonian introduced by Schr\"{o}dinger [30], and later
developed by Infeld and Hull [31], It is useful to discover for
different potentials with equivalent energy spectra in
non-relativistic quantum mechanics. Various aspects has been
studied within the formalism of SUSYQM [32]. This paper is
arranged as follows: In Sec. II we introduce the Hamiltonian
Hierarchy method. In Sec. III we apply the method to solve the
Schr\"{o}dinger equation with PT-symmetric and non-PT-symmetric
non-Hermitian forms of the q-deformed Morse potential. In Sec. IV
we discuss the results.

\section{SUSYQM and Hamiltonian Hierarchy Method}
\noindent Supersymmetric algebra allows us to write Hamiltonians
as [30]

\begin{equation}
H_{\pm}=-\frac{\hbar^{2}}{2m}\frac{d^{2}}{dx^{2}}+
V_{\pm}(x),
\end{equation}

\noindent where
The supersymmetric partner potentials $V_{\pm}(x)$
in terms of the superpotential $W(x)$ are given by

\begin{equation}
V_{\pm}(x)=W^{2}\pm\frac{\hbar}{\sqrt{2m}}
\frac{dW}{dx}.
\end{equation}

\noindent
The superpotential has a definition

\begin{equation}
W(x)=-\frac{\hbar}{\sqrt{2m}}\left[\frac{d\ln\Psi_{0}^{(0)}(x)}{dx}\right],
\end{equation}

\noindent
where, $\Psi_{0}^{(0)}(x)$ denotes the ground state wave
function that satisfies the relation

\beq
\Psi_{0}^{(0)}(x)=N_{0}~\exp\left[-\frac{\sqrt{2m}}{\hbar}
\int^{x} W(x^{\prime})~dx^{\prime}\right].
\eeq

\noindent
The Hamiltonian $H_{\pm}$ can also be written in terms of the
bosonic operators $A^{-}$ and $A^{+}$

\begin{equation}
H_{\pm}=A^{\mp}~A^{\pm},
\end{equation}

\noindent
where

\begin{equation}
A^{\pm}=\pm\frac{\hbar}{\sqrt{2m}}\frac{d}{dx}+W(x).
\end{equation}

It is remarkable result that the energy eigenvalues of $H_{-}$ and
$H_{+}$ are identical except for the ground state. In the case of
unbroken supersymmetry, the ground state energy of the Hamiltonian
$H_{-}$ is zero $\left(E_{0}^{(0)}=0\right)$ [30]. In the
factorization of the Hamiltonian, the Eqs. (1), (5) and (6) are
used respectively. Hence, we obtain

\begin{eqnarray}
H_{1}(x)&=&-\frac{{\hbar}^{2}}{2m}\frac{d^{2}}{dx^{2}}+V_{1}(x)\nonumber\\[0.2cm]
        &=&(A_{1}^{+}~A_{1}^{-})+E_{1}^{(0)}.
\end{eqnarray}

\smallskip
\noindent
Comparing each side of the Eq. (7) term by term, we get
the Riccati equation for the superpotential $W_{1}\left(x\right)$

\begin{equation}
W_{1}^{2}-W_{1}^{'}=\frac{2m}{{\hbar}^{2}}\left(V_{1}(x)-E_{1}^{(0)}\right).
\end{equation}

\smallskip
Let us now construct the supersymmetric partner
Hamiltonian $H_{2}$ as

\begin{eqnarray}
H_{2}(x)&=&-\frac{{\hbar}^{2}}{2m}\frac{d^{2}}{dx^{2}}+V_{2}(x)\nonumber\\[0.2cm]
        &=&\left(A_{2}^{-}~A_{2}^{+}\right)+E_{2}^{(0)},
\end{eqnarray}

\noindent
and Riccati equation takes
\begin{equation}
W_{2}^{2}+W_{2}^{'}=\frac{2m}{{\hbar}^{2}}\left(V_{2}(x)-E_{2}^{(0)}\right).
\end{equation}

\smallskip
\noindent Similarly, one can write in general the  Riccati
equation and Hamiltonians by iteration as

\smallskip
\begin{eqnarray}
W_{n}^{2}\pm
W_{n}^{'}&=&\frac{2m}{{\hbar}^{2}}\left(V_{n}(x)-E_{n}^{(0)}\right)\nonumber\\[0.2cm]
         &=&\left(A_{n}^{\pm}~A_{n}^{\mp}\right)+E_{n}^{(0)},
\end{eqnarray}

\noindent
and

\begin{eqnarray}
H_{n}(x)&=&-\frac{{\hbar}^{2}}{2m}\frac{d^{2}}{dx^{2}}+V_{n}(x)\nonumber\\[0.2cm]
        &=& A_{n}^{+}~A_{n}^{-}+E_{n}^{(0)},\quad\quad n=1,2,3,\ldots
\end{eqnarray}

\noindent where \begin{equation}
A_{n}^{\pm}=\pm\frac{\hbar}{\sqrt{2m}}\frac{d}{dx}+\frac{d}
{dx}\left(\ln\Psi_{n}^{(0)}(x)\right).
\end{equation}

\smallskip
\noindent Because of the SUSY unbroken case, the partner
Hamiltonians satisfy the following expressions [30]

\begin{equation}
E_{n+1}^{(0)}=E_{n}^{(1)}, \quad with \quad E_{0}^{(0)}=0,\quad
n=0,1,2,\ldots
\end{equation}

\smallskip
\noindent and also the wave function with the same eigenvalue can
be written as [30]

\begin{equation}
\Psi_{n}^{(1)}=\frac{A^{-}~\Psi_{n+1}^{(0)}}{\sqrt {E_{n}^{(0)}}},
\end{equation}

\noindent
with

\begin{equation}
\Psi_{n+1}^{(0)}=\frac{A^{+}~\Psi_{n}^{(1)}}{\sqrt {E_{n}^{(0)}}}.
\end{equation}

\smallskip
This procedure is known as the hierarchy of Hamiltonians.

\section{Calculations}
\subsection{The General q-deformed Morse case}
\noindent Let us first consider the generalized Morse potential as
[16]
\begin{equation}
V_{M}(x)=V_{1}~e^{-2~a~ x}-V_{2}~e^{-a~x}.
\end{equation}

\noindent where $V_{1}$ and $V_{2}$ are real parameters. By
comparing the Eq. (17) with the following equation

\begin{equation}
V_{M}(x)=D(e^{-2~a~x}-2qe^{-a~x}),
\end{equation}

\noindent we have $V_{1}=D$ and $V_{2}=2qD$. Therefore we can
construct the hierarchy of Hamiltonians for Schr\"{o}dinger
equation with $\ell=0$,

\begin{equation}
[-\frac{d^{2}\Psi}{dx^{2}}+\mu^{2}(e^{-2~a~x}-2qe^{-a~x})]\Psi(x)=\varepsilon\Psi(x),
\end{equation}

\smallskip
\noindent where $\ds{\mu^{2}=\frac{2mV_{1}}{a^{2}\hbar^{2}}}$ and
$\ds{E=\varepsilon\frac{a^{2}\hbar^{2}}{2m}}$.
\smallskip
\noindent We can also write the Riccati equation as
\begin{equation}
W_{1}^{2}-W_{1}^{\prime}+\varepsilon_{0}^{(1)}=V_{1}(x).
\end{equation}

\noindent Here $V_{1}(x)$ is the superpartner of the
superpotential $W_{1}(x)$. Following by ansatz equation, we have

\begin{equation}
W_{1}(x)=-\mu~e^{-a~x}+q~\delta,
\end{equation}

\noindent and inserting this into the Eq. (20), we get
\begin{equation}
\delta=(\mu-\frac{a}{2q}),
\end{equation}

\noindent with the first ground state energy
\smallskip
\begin{equation}
\varepsilon_{0}^{(1)}=-q^{2}(\mu-\frac{a}{2q})^{2}.
\end{equation}

\noindent In order to construct the other superpartner potential
$V_{2}(x)$, we will solve the equation

\begin{equation}
W_{1}^{2}+W_{1}^{\prime}+\varepsilon_{0}^{(1)}=V_{2}(x).
\end{equation}

\noindent Then we can find the second member superpotential as

\begin{equation}
W_{2}(x)=-\mu~e^{-a~x}+q~\kappa.
\end{equation}

\noindent Now, putting this ansatz into the Eq. (20), we get

\smallskip
\begin{equation}
\kappa=(\mu-\frac{3a}{2q}),
\end{equation}

\smallskip
\noindent with

\smallskip
\begin{equation}
\varepsilon_{0}^{(2)}=-q^{2}(\mu-\frac{3a}{2q})^{2}.
\end{equation}

\smallskip
\noindent By similar iterations, one can get the general results

\begin{equation}
W_{n+1}(x)=-\mu~e^{-ax}+q\left[\mu-\frac{a}{q}(n+\frac{1}{2})\right],
\end{equation}

\begin{equation}
V_{n+1}(x)=\mu^{2}(e^{-2ax}-2qe^{-ax})+2na\mu~e^{-ax},
\end{equation}

\begin{equation}
E_{n+1}^{(\ell=0)}=-q^{2}\left[\mu-\frac{a}{q}(n+\frac{1}{2})\right]^{2},
\end{equation}

\noindent and ground state wave function
\begin{equation}
\Psi_{0}(x)=N~\exp\{-\widetilde{\mu}e^{-ax}+q\left[\mu
-\frac{a}{q}(n+\frac{1}{2})\right]x\}.
\end{equation}

\noindent where we choose $\widetilde{\mu}=a\mu$ and set $(\hbar=2m=1)$
in Eq. (30).

\subsection{Non-PT-symmetric and Non-Hermitian Morse case}
\noindent Let us now consider the Eq. (17) with respect to
$V_{1}\rightarrow~D$ as real and $V_{2}\rightarrow~2iqD$ as complex
parameters. Hence we construct the hierarchy of Hamiltonian of the
Schr\"{o}dinger equation for the complexified Morse potential as

\begin{equation}
[-\frac{d^{2}\Psi}{dx^{2}}+\mu^{2}(e^{-2~a~x}-2iqe^{-a~x})]\Psi(x)=\varepsilon\Psi(x),
\end{equation}

\smallskip
\noindent where
$\ds{\mu^{2}=\frac{2mV_{1}}{a^{2}\hbar^{2}}}$ and
$\ds{E=\varepsilon\frac{a^{2}\hbar^{2}}{2m}}$.

\smallskip
\noindent Applying the hierarchy of Hamiltonians as in the
previous section, the $(n+1)$-th member results will be

\begin{equation}
W_{n+1}(x)=-\mu~e^{-a~x}+q\left[i\mu-\frac{a}{q}(n+\frac{1}{2})\right],
\end{equation}

\begin{equation}
V_{n+1}(x)=\mu^{2}(e^{-2ax}-2qe^{-ax})+2na\mu~e^{-ax},
\end{equation}

\begin{equation}
E_{n+1}^{(\ell=0)}=-q^{2}\left[i\mu-\frac{a}{q}(n+\frac{1}{2})\right]^{2},
\end{equation}

\noindent with
\begin{equation}
\Psi_{0}(x)=N~\exp~\{-\widetilde{\mu}e^{-ax}+q\left[i\mu-\frac{a}{q}(n
+\frac{1}{2})\right]x\},
\end{equation}
\noindent where  $\widetilde{\mu}=a\mu$.

\subsection{PT-symmetric and Non-Hermitian Morse Case}
Let us assume the potential parameters
$V_{1}=(\alpha+i\beta)^{2}$ and $V_{2}=(2\gamma+1)(\alpha+i\beta)$
in Eq. (17). Here we choose $\alpha$ and $\beta$ and
$\gamma=-\frac{1}{2}+q(\alpha+i\beta)$. When $a\rightarrow~i~a$ in
Eq. (17) and choosing $V_{1}$ and $V_{2}$ as in the previous section, the
potential form will be

\begin{equation}
V_{M}(x)=(\alpha+i\beta)^{2}(e^{-2iax}-2qe^{-iax}).
\end{equation}

\noindent The ansatz equation is
\begin{equation}
W_{1}(x)=\xi~e^{-ia~x}+iq~\delta.
\end{equation}

\smallskip
\noindent As a result the Schr\"{o}dinger equation can be written by using
the Eq. (19) for
$\ds{\mu^{2}=\frac{2m(\alpha+i\beta)^{2}}{a^{2}\hbar^{2}}}$ and
$\ds{E=\varepsilon\frac{a^{2}\hbar^{2}}{2m}}~(if E<0)$. Applying the same
procedure again, one can get

\begin{equation}
W_{n+1}(x)=\xi~e^{-ia~x}+iq\left[i\xi-\frac{a}{q}(n+\frac{1}{2})\right],
\end{equation}

\begin{equation}
V_{n+1}(x)=\mu^{2}(e^{-2iax}-2qe^{-iax})+2in\xi~a~e^{-iax},
\end{equation}

\begin{equation}
E_{n+1}^{(\ell=0)}=-q^{2}\left[i\xi-\frac{a}{q}(n+\frac{1}{2})\right]^{2},
\end{equation}

\noindent with
\begin{equation}
\Psi_{0}(x)=N~\exp~\{-\widetilde{\xi}e^{-iax}+iq\left[i\xi
-\frac{a}{q}(n+\frac{1}{2})\right]x\},
\end{equation}
\noindent where  $\widetilde{\mu}=ia\mu$.

\section{Conclusions}
\noindent We have used the PT-symmetric formulation developed recently 
within non-relativistic quantum mechanics to a more general Morse 
potential. We have solved the Schr\"{o}dinger equation in one dimension by 
applying Hamiltonian hierarchy method within the framework of SUSYQM. We 
discussed many different complex forms of this potential. Energy 
eigenvalues and 
corresponding eigenfunctions are obtained exactly. We also point out that 
the exact results obtained for the complexified Morse potential may 
increase the number of interesting applications in the study of different 
quantum mechanical systems. In the case of $\beta=0$ in Eq. (41), there is 
only real spectra, when 
$\alpha=0$, otherwise there exists a complex-valued energy
spectra. This implies that broken PT-symmetry doesn't occur spontaneously.
Moreover, PT-/non-PT-symmetric non-Hermitian solutions have  the same 
spectra. We also note that both real and imaginary part of the energy 
eigenvalues corresponds to the anharmonic and harmonic oscillator 
solutions. The $(n+1)-th$ member
superpotential, its superpartner and also  corresponding ground
state eigenfunctions of PT-symmetric non-Hermitian potentials
satisfy the condition of PT-invariance though the others are not.

\end{document}